\documentclass[aps,prl,twocolumn,superscriptaddress,floatfix]{revtex4-1}
\usepackage[dvips]{graphicx}
\usepackage{dcolumn}
\usepackage{bm}

\begin{document}

\title{Real-time observation of discrete Andreev tunneling events}

\author{V. F. Maisi}
\email{ville.maisi@mikes.fi}
\affiliation{Low Temperature Laboratory, Aalto University, P.O.~Box 13500, 00076 Aalto, Finland}
\affiliation{Centre for Metrology and Accreditation (MIKES), P.O. Box 9, 02151 Espoo, Finland}

\author{O.-P. Saira}
\affiliation{Low Temperature Laboratory, Aalto University, P.O.~Box 13500, 00076 Aalto, Finland}

\author{Yu. A. Pashkin}
\altaffiliation[on leave from ]{Lebedev Physical Institute, Moscow 119991, Russia} \affiliation{NEC Green
Innovation Research Laboratories and RIKEN Advanced Science Institute, 34 Miyukigaoka, Tsukuba, Ibaraki
305-8501, Japan}

\author{J. S. Tsai}
\affiliation{NEC Green Innovation Research Laboratories and RIKEN Advanced Science Institute, 34 Miyukigaoka,
Tsukuba, Ibaraki 305-8501, Japan}

\author{D. V.~Averin}
\affiliation{Department of Physics and Astronomy, Stony Brook University, SUNY, Stony Brook, NY 11794-3800, USA}

\author{J. P. Pekola}
\affiliation{Low Temperature Laboratory, Aalto University, P.O.~Box 13500, 00076 Aalto, Finland}

\begin{abstract} 
We provide a direct proof of two-electron Andreev transitions in a 
superconductor - normal metal tunnel junction by detecting them in a 
real-time electron counting experiment. Our 
results are consistent with ballistic Andreev transport with an order of magnitude 
higher rate than expected for a uniform barrier, suggesting that only 
part of the interface is effectively contributing to the transport. 
These findings are quantitatively supported by our direct current 
measurements in single-electron transistors with similar tunnel barriers.

\end{abstract}

\maketitle
Electronic transport across a boundary between conductors with dissimilar carriers is a non-trivial process. Of particular interest in this respect is the transport through a superconductor - normal metal interface that at low energies is dominated by Andreev reflection~\cite{andreev64,btk,eiles93,lafarge93,hekking94,pothier94,sukumar08,vasenko10,wei10,greibe10}, where a Cooper pair in a superconductor is converted into two electrons in the normal metal or vice versa. Here we employ electron counting techniques~\cite{td1,td2,td3,td4,td5,td6,saira10} to detect these Andreev events. Since the observed rate depends on the 
coherence of the two electrons involved in the transition, we obtain, as a result, a fingerprint of the junction electrodes and the tunnel barrier. 

\begin{figure}[t]
	\centering
	\includegraphics[width=0.49\textwidth]{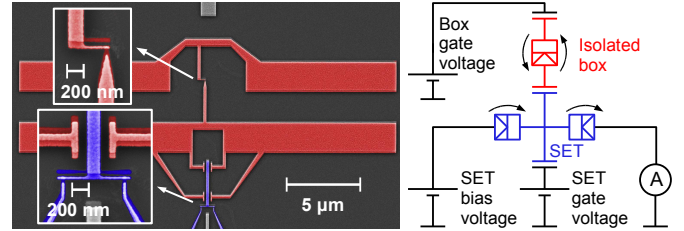}
	\caption{\label{fig:sample} (color online) Scanning electron micrograph of the measured structure and the schematic layout of the measurement set-up. The isolated electron box consists of two metallic islands, seen as $25\ \mathrm{\mu m}$ long rectangles (colored red). They are connected to each other by a normal metal - insulator - superconductor (NIS) tunnel junction. Tunneling of electrons through the junction is monitored with a DC SET electrometer (in blue) coupled capacitively to one of the box islands. It is voltage biased, and the gate voltage of the SET sets the electrometer to a charge sensitive point, and the gate voltage of the box tunes the energy levels of the charge states in it. The NIS junction (top) and detector (bottom) are shown magnified on the left side of the main micrograph.}
\end{figure}

The techniques used for observing individual electrons are based on the Coulomb blockade effect where the electrostatic energy of a small metallic island changes noticeably when only one elementary charge $e$ is placed on or removed from it. 
In the present experiment, we employ an isolated single-electron box where a superconducting island is connected to a normal metal one~\cite{saira10}, but neither of these two is connected galvanically to the external circuitry.  
 The electron tunneling rates between the islands are then sufficiently low to be monitored by low-frequency electrometry and are described in detail by relatively simple theoretical considerations \cite{saira10,pekola10}. We use a single-electron transistor (SET) \cite{td1,td2,td3,td4,td5,sac,al,fd,eNoise1,saira10} as an ultrasensitive electrometer. With charge sensitivity as good as $10^{-5}e/\sqrt{\rm Hz}$ \cite{eNoise1,krupenin98,roschier01}, it is capable to detect individual electrons with high precision.
In Fig. \ref{fig:sample}, we show a micrograph of our sample fabricated by standard e-beam processing.

\begin{figure*}[t]
	\centering
	\includegraphics[width=0.9\textwidth]{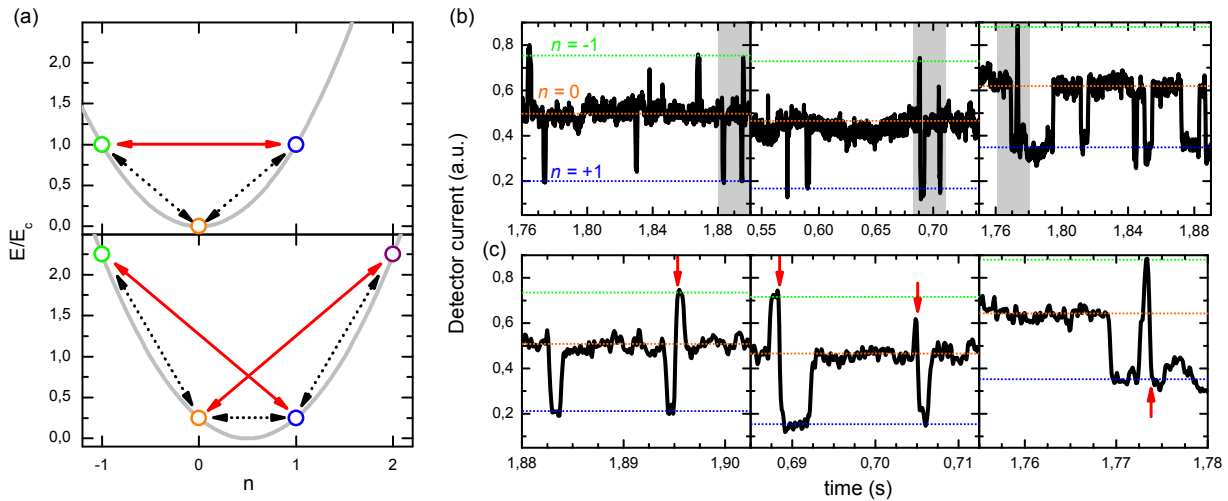}
	\caption{\label{fig:traces} (color online) Energy levels of the various charge states and typical observed time traces of the current through the detector. (a) Low lying levels of the box in the Coulomb blockade (upper panel) and at degeneracy (lower panel). Dotted (black) and solid (red) arrows indicate one- and two-electron processes, respectively. (b) Measured time traces of the detector current showing the charge state of the electron box as a function of time. The leftmost panel presents the case of Coulomb blockade. The rightmost panel depicts the opposite limit where the two charge states are equal in energy (degeneracy). The trace in the centre is taken halfway between these two cases. (c) Grey sections of the traces of (b) zoomed. Vertical arrows indicate two-electron events.}
\end{figure*}

The tunneling rates and resulting charge distribution between the two islands of the isolated box can be adjusted with an offset charge induced by a gate voltage. The electrostatic energy of a state with $n$ excess electrons on one of the islands is given by  $E_n = E_c(n-n_g)^2$, where $E_c$ is the charging energy for individual electrons, and $n_g$ is the normalized offset charge that can be viewed as the polarization charge on the gate capacitor and determines the energetically preferred way to occupy the different charge states $n$~(Ref.~\cite{al}). In Fig.~\ref{fig:traces} (a), the two extreme cases are shown. In the Coulomb blockade regime for single electrons, $n_g$ is an integer and the state $n = n_g$ has the minimal energy $E_n = 0$. To enter an excited state, one electron can either tunnel into or out of the island  (dotted black lines with arrows in  Fig.~\ref{fig:traces} (a)) but energy $E_c$ has to be provided for the tunneling electron in addition to the Cooper pair breaking energy equal to or larger than the superconducting energy gap $ \Delta$~\cite{giaever60}. In the other extreme, at degeneracy with half-integer $n_g$, two electron states differing by charge $e$ have equal minimal energy and hence are equally populated. The tunneling rate between them is higher than in the Coulomb blockade regime as no extra energy for charging is needed. For Andreev reflection (solid red lines with arrows in Fig.~\ref{fig:traces} (a)), the energy cost of charging is calculated similarly, but now the initial and final states are separated by two electrons, and the energy cost of breaking a Cooper pair is avoided since complete pairs tunnel at once.

In the experiment (see supplemental material for additional details~\cite{suppmat}), we measured time traces of the detector current at various biasing conditions of the box. With all the other parameters fixed, the detector gate voltage was adjusted to maximize charge sensitivity and dynamic range. The observed current jumps (Fig. 2b and c) are attributed to the tunneling events between the two islands. The switching rate depends on the gate voltage of the box and hence on its charge state, being lowest in the Coulomb blockade regime (leftmost panel) and highest at degeneracy (rightmost panel).
The events observed in the traces (grey regions of Fig.~\ref{fig:traces} (b) zoomed in \ref{fig:traces} (c)), indicate that individual electrons tunnel between the islands: In the Coulomb blockade regime they hop rarely from the lowest ($n=0$) to the higher ($n=\pm1$) energy states and back, while at degeneracy the electrons tunnel frequently between the two lowest states and only occasionally the system enters a higher lying level. More interestingly, the traces show also the coincident events, pointed out by the vertical arrows, where two electrons appear to tunnel simultaneously. In the following, we show that most of these events represent Andreev tunneling.

\begin{figure}[t]
	\centering
	\includegraphics[width=0.45\textwidth]{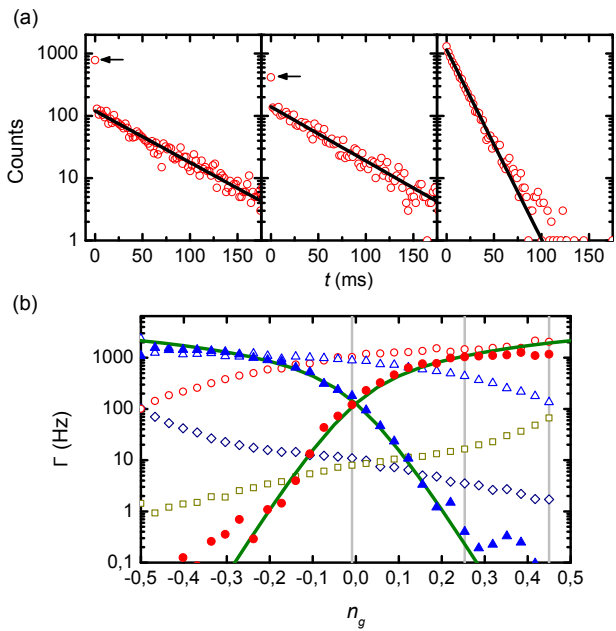}
	\caption{\label{fig:lifetimes} (color online) Lifetime distributions and tunneling rates. (a) Lifetime distributions of the $n = 0$ charge state in the Coulomb blockade, at an intermediate gate position and close to degeneracy, from left to right. The corresponding gate positions are indicated by vertical lines in (b). The distribution is exponential as expected for a Poisson process. However, at small lifetimes an anomaly, pointed out with the horizontal arrows, is observed as two-electron events are being counted as fast events. The anomaly vanishes as one approaches degeneracy since the energy cost is high for the two-electron events. (b) One-electron (open symbols) and two-electron (solid symbols) tunneling rates estimated from the counted events at the base temperature. Red circles and blue triangles denote the tunneling rates out of states $n = -1$ and $1$. Dark blue diamonds and dark yellow squares denote tunneling rates out of state $0$ to states $-1$ and $1$. Dark green solid lines present calculated Andreev rates with $\zeta = 4 \cdot 10^{-5}$. }
\end{figure}

Due to the finite measurement bandwidth, limited to $1\ \mathrm{kHz}$ by the dc readout of the electrometer, events resembling the two-electron Andreev tunneling could in principle arise from almost coincidental tunneling of two independent quasiparticles. To assess this option, we recorded time traces for several minutes at each gate offset value. From the traces we determined the distribution of the time $t$ spent in the state $n = 0$ before a transition took place. In Fig.~\ref{fig:lifetimes} (a) we show such a distribution on the left for Coulomb blockade ($n_g=0$), in the centre for $n_g=0.25$, and on the right near degeneracy ($n_g=0.45$). A direct transition (Andreev tunneling) between states $n = \pm 1$ contributes here as essentially a $t=0$ event since the time separation between the two electrons tunneling in the Andreev process should be on the order of $\hbar/\Delta$, which is many orders of magnitude smaller than the time scales relevant in Fig~\ref{fig:lifetimes} (a).   Overall, the distribution is exponential as we have Poisson distributed one-electron tunneling processes. However, at small lifetimes in the Coulomb blockade regime, the data point indicated by the horizontal arrow does not follow the exponential dependence and corresponds to excessively many events. This  clear separation of the short-lifetime events from the one-electron transitions shows that the majority of these events are not coincidental one-electron tunneling but rather two electrons tunneling concurrently. When the box offset charge is adjusted closer to degeneracy, the anomalous datapoint gradually merges to the rest of the data, in accordance with its interpretation in terms of Andreev transitions, since the energy cost of the two-electron tunneling increases. Eventually at degeneracy, almost no two-electron events are observed. We emphasize that the charging energy should be small, $E_c < \Delta$, for Andreev tunneling to occur, since for large $E_c$, it is not energetically favourable~\cite{saira10}.

\begin{figure}[!ht]
	\centering
	\includegraphics[width=0.44\textwidth]{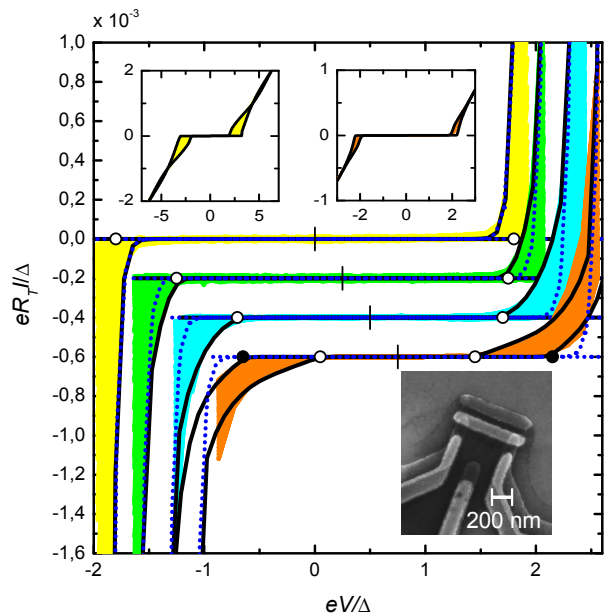}
	\caption{\label{fig:ivs} (color online) Detection of Andreev current in SETs. Colored regions show measured subgap current for devices with charging energies $E_c/k_B = 2.3$, $1.9$, $1.5$ and $0.86\ \mathrm{K}$ from top (yellow) to bottom (orange) at all gate offset values. The insets depict the larger scale measurements for samples with the highest and lowest charging energies and a scanning electron micrograph of one of the measured SETs. Solid black lines are fitted theoretical curves at degeneracy (maximal current) and in Coulomb blockade (minimal current) when Andreev tunneling is taken into account. Dotted blue lines present fits excluding Andreev processes. Tunneling resistances of the samples were $R_T = 129$, $78$, $55$ and $31\ \mathrm{k\Omega}$ in the order of decreasing $E_c$, and superconducting gap $\Delta = 216\ \mathrm{\mu eV}$ for all of them. Fitting parameter for the magnitude of the Andreev processes was $\zeta \approx 4 \cdot 10^{-4}$. (Note that to increase the measured signal the SET junctions were oxidized less to have larger specific conductance than the box junction.) Open (solid) circles present the expected thresholds,  $eV = \pm 2 E_c$, ($eV = \pm 4 E_c$)  for Andreev tunneling at degeneracy (Coulomb blockade). }
\end{figure}

For quantitative analysis we counted the number of one- and two-electron events, and normalized those by the time spent in the initial state to obtain the corresponding tunneling rate. The results for both processes at the temperature of 60 mK are shown in Fig.~\ref{fig:lifetimes} (b). The parameter values which determine the first order tunneling rates in the box junction, the tunneling resistance $R_T = 2\ \mathrm{M \Omega}$, superconducting energy gap $\Delta = 216\ \mathrm{\mu eV}$, and the charging energy $E_c = 0.2\ \Delta$, were determined from the one-electron tunneling rates measured at temperatures ranging from $60\ \mathrm{mK}$ to $200\ \mathrm{mK}$ as in Ref.~\cite{saira10}.

To calculate the two-electron rates, we assume ballistic model of the tunnel barrier, which fits well our experiments with small junctions. Apart from the parameter values obtained from one-electron tunneling, the only parameter to be determined for the two-electron tunneling is its overall magnitude controlled by $\zeta \equiv (\hbar /R_Te^2)/{\mathcal N}$, where ${\mathcal N}$ is the effective number of the conduction channels in the junction, ${\mathcal N}=A/A_{ch}$ \cite{averin08}. The junction area $A = 40\ \mathrm{nm\ x}\ 35\ \mathrm{nm}$ is estimated from the scanning electron micrograph of the tunnel junction, and the area of one conduction channel $A_{ch} = 30\ \mathrm{nm^2}$ is determined independently from measurements of the SET current described in detail below. The results of the calculation of the Andreev rates with these parameter values are plotted  as solid lines in Fig.~\ref{fig:lifetimes} (b), and show good agreement with the measured rates. For completely uniform tunnel junction, however, the area of a conduction channel should be about $A_{ch} \simeq 2\ \mathrm{nm^2}$, see supplemental material~\cite{suppmat}. We note that the measured value is about one order of magnitude higher than the theoretical estimate, which agrees with the conclusions in, e.g. \cite{pothier94,greibe10}, for larger tunnel junctions that exhibit diffusive Andreev tunneling. We attribute this to the imperfections of the barrier, with only a small portion of the junction area dominating the transport. This fact can be understood by noting that the standard parameters of the aluminum oxide tunnel barriers~\cite{suppmat,likh_08,barr} imply that, for example, variation of the barrier thickness by one atomic layer (i.e. by about 0.3 nm) results in the specific conductance change by about a factor of 10. This means that even relatively small fluctuations of the barrier thickness of the magnitude of one to two atomic layers in comparison with the average barrier thickness of $6$ atomic layers can reduce the effective area of the region dominating the barrier transparency to about 10\% of the total junction area.

As an additional way of extracting parameters of the two-electron tunneling, we measured current-voltage characteristics of basic SET structures with superconducting leads and a normal metal island. A micrograph of one of the devices is shown in the inset of Fig.~\ref{fig:ivs} (a). All four SETs of Fig.~\ref{fig:ivs} were fabricated in the same batch so that only the junction area ($E_c$) varies. Here Andreev tunneling has to be extracted from the measured data containing a contribution also from one electron processes. The SET was biased at voltage $V$ and the current $I$ flowing through the device was measured. At each bias voltage, the gate offset of the island was varied so that the maximal and minimal current, corresponding to degeneracy and Coulomb blockade, respectively, were observed. The measured current values between these extremes are shown as the colored regions in Fig.~\ref{fig:ivs} (b) for four devices having different charging energies. For the highest charging energy (top), we have no current flowing in the subgap region $eV < 2\Delta$, apart from the thermally activated single-electron tunneling. When the charging energy is lowered, the onset of the Andreev current penetrates into the subgap region with the threshold at $eV=2E_c$, seen both in the experimental data and in the fits including Andreev processes (solid black lines), as the energy cost for two electrons tunneling into the island is lowered. With the high charging energies, $E_c \geq \Delta$, it is sufficient to take into account only one-electron tunneling in the fits (dotted blue lines). On the large current scale, the simulations with and without Andreev current are indistinguishable as shown in the insets for samples with the highest and the lowest charging energy. In the numerical simulations for all the samples, we obtain as the only fit parameter the area of one conduction channel to be $A_{\mathrm{ch}} = 30\ \mathrm{nm^2}$. This value was adopted to the analysis of the counting data above.

In conclusion, we have detected in time-domain individual Andreev transitions allowing us to distinguish higher order tunneling from the usual one-electron processes. These counting experiments are consistent with our direct current measurements, and they both indicate that the tunnel barriers are non-uniform.

The work has been supported partially by the Academy of Finland, Technology Industries of Finland
Centennial Foundation, V\"ais\"al\"a Foundation, the Finnish National Graduate School in Nanoscience, MEXT kakenhi "Quantum Cybernetics", the JSPS through its FIRST Program and the European Community's Seventh Framework Programme under Grant Agreement No. 218783 (SCOPE). D.V.A. was supported in part by IARPA.

\bibliography{prlbib}

\end{document}